\documentclass[prd,aps,amssymb,showpacs]{revtex4}
\usepackage{amsmath,amsfonts}

\begin{document}

\title{Renormalization for self-potential of a scalar charge in static space-times}

\author{Arkady A. Popov}
\email{apopov@ksu.ru}
\affiliation{
Department of Mathematics, Tatar State University of Humanities and Education, Tatarstan 2, Kazan 420021, Russia\\
Department of Physics, Kazan State University, Kremlevskaya 18, Kazan, 420008, Russia}

\begin{abstract}

A method is presented which allows for the renormalization
of the self-potential for a scalar point charge at rest in static
curved space-time. The method is suitable for the scalar field with
arbitrary mass $m$ and coupling to the scalar curvature. The asymptotic
behavior of self-potential is obtained in the limit in which the Compton
wave-length $1/m$ of the massive scalar field is much smaller than
the characteristic scale of curvature of the background gravitational field.
The self-force is calculated in this limit.

\end{abstract}

\maketitle

\section{Introduction}

It is known that a charged particle interacts with the field, the source
of which is this particle.  In flat space-time the effect
is determined by the derivative of acceleration of the charge \cite{Dirac:1938}.
The origin of self-interaction in curved space-times is associated
with the nonlocal structure of the field.
In static curved space-times and space-times with nontrivial topology the self-force
can be nonzero even for the charge at rest (here and below the words ''at rest''
mean that the velocity of charge is collinear to the timelike Killing vector
which always exists in a static space-time). The formal expression for
the electromagnetic self-force in an arbitrary curved space-time has been derived
by DeWitt and Brehme \cite{DeWBre60} and a correction was later provided
by Hobbs \cite{Hobbs:1968}. Mino, Sasaki, and Tanaka \cite{MST:1997} and independently
Quinn and Wald \cite{Quinn:1997} obtained a similar expression for the gravitational
self-force on a point mass. The self-force on a charge interacting
with a massless minimally coupled scalar field was considered by Quinn
\cite{Quinn:2000}. A discussion of the self-force in detail may be found in reviews
\cite{Poi04,Detw05,Khu05}.

Calculating the self-force one must evaluate the field that the point charge induces
at the position of the charge. This field diverges and must be renormalized.
There are different methods of such type of renormalization. Some of them are
reviewed in \cite{Hikida:2005,Barack:2009}. Note also the zeta function method
\cite{Lousto:2000} and the ''massive field approach'' for the calculation of the self-force
\cite{1Roth:2004,2Roth:2004}. In the ultrastatic space-times the renormalization
of the field of static charge can be realized by the subtraction of the first
terms from DeWitt-Schwinger asymptotic expansion of a three-dimensional
Euclidean Green's function \cite{Khus07,Kra08,BezKhu09,Popov:2010}.
In this paper a similar approach expands to the case of static space-times.
In the framework of the suggested procedure one subtracts some
terms of expansion of the corresponding Green's function of a massive scalar
field with arbitrary coupling to the scalar curvature from
the divergent expression obtained. The quantities of
terms to be subtracted are defined by a simple rule -- they no longer
vanish as the field's mass  goes to infinity.
Such an approach is similar to renormalization introduced in the context
of the quantum field theory in curved space-time \cite{DW,Chr:1978}.
The Bunch and Parker method \cite{BP:1979} is used for
expansion of the corresponding Green's function of a scalar field.

The organization of this paper is as follows. In Sec.
\ref{Sec:Renormalization},  we expand the potential of a scalar point
charge as two points (in 3D space) function in powers of distance
between these points and determine the procedure of renormalization.
In Sec. \ref{Sec:BlackHole} we calculate the renormalized self-potential
of the scalar point charge in Schwarzschild space-time, as an example of
the presented method.
We discuss the results in Sec.
\ref{Sec:Conclusion}. Our conventions are those of Misner, Thorne,
and Wheeler \cite{MTW:73}. Throughout this paper, we use units $c =
G = 1$.

\section{Renormalization}\label{Sec:Renormalization}

Let us consider an equation for the scalar massive field with source
\begin{equation} \label{meq}
{\phi_m}^{;\mu}_{;\mu} - \left(m^2+\xi R \right) \phi_m = -J = -4\pi q \int
\delta^{(4)}(x-x_0(\tau)) \frac{d\tau}{ \sqrt{-g^{(4)}}},
\end{equation}
where $\xi$ is a coupling of the scalar field with mass $m$ to the scalar
curvature $R$, $g^{(4)}$ is the determinant of the metric
$g_{\mu \nu}$, $q$ is the scalar charge and $\tau$ is its proper time.
The world line of the charge is given by ${x_0^\mu} (\tau)$.
The metric of static space-time can be presented as follows:
\begin{equation} \label{metric}
ds^2= -g_{tt}(x^i)d t^2+g_{j k}(x^i)d x^j d x^k,
\end{equation}
where $i, j, k = 1, 2, 3$. This means that one can write the field
equation in the following way:
        \begin{eqnarray} \label{beq0}
        \frac{1}{\sqrt{g_{tt}}\sqrt{g^{(3)}}} \frac{\partial}{\partial x^j}
        \left( \sqrt{g_{tt}}\sqrt{g^{(3)}} g^{j k} \frac{\partial
        \phi_m(x^i; x_0^i)}{\partial x^k}   \right)
        -\left(m^2+\xi R(x)\frac{}{} \right) \phi_m(x^i; x_0^i) \nonumber \\
        =-\frac{4 \pi q \delta^{(3)}(x^i, x_0^i)}{\sqrt{g^{(3)}}},
        \end{eqnarray}
where $m$ is the mass of the scalar field,
$g^{(3)}=\det g_{i j}$  and we take into account that $d\tau/dt=\sqrt{g_{tt}}$
for the particle at rest.
In the case
\begin{equation}
m \gg 1 / L,
\end{equation}
where $L$ is the characteristic curvature scale of the background geometry,
it is possible to construct the iterative procedure of the solution of
Eq. (\ref{beq0}) with small parameter $1/(m L)$ \cite{DW,Chr:1978,BP:1979}.
This expansion can be used in the regularization procedure of Rosenthal \cite{1Roth:2004}
\begin{equation}\label{Rosenthal}
f_\mu^{self}(x_0)=q\lim_{m \rightarrow \infty}\left\{
\lim_{\delta \rightarrow 0} \frac{\displaystyle
\partial \left( \frac{}{} \phi(x;x_0)-\phi_m(x;x_0)\right)}{\partial x^\mu}+
\frac{q m^2 n_\mu(x_0)}{2}+\frac{q m a_\mu(x_0)}{2}\right\},
\end{equation}
because this procedure demands the calculation of the expansion of $\phi_m(x;x_0)$ in terms
of $x^\mu-x^\mu_0$ and $1/m$ accurate to order
$O\left( \frac{}{} (x-x_0)^2 \right)+O\left( 1/m \right)$ only.
In the expression (\ref{Rosenthal}) $\phi(x;x_0)$ is the massless
field induced by scalar charge $q$, and $x$ is a point near the charge's
world line $x_0(\tau)$, defined as follows.
At $x_0$ we construct a unit spatial vector $n^\mu$,
which is perpendicular to the object's world line but is otherwise
arbitrary (i.e. at $x_0$ we have $n^\mu n_\mu=1$ , $n^\mu
u_\mu=0$). In the direction of this vector we construct a
geodesic, which extends out an invariant length $\delta$ to the
point $x(x_0,n^\mu,\delta)$; throughout this paper $u^\mu$
and $a^\mu$ denote the object's four-velocity and four-acceleration,
at $x_0$, respectively.

To construct the expansion of $\phi_m(x;x_0)$,
let us consider the equation for the three-dimensional
Green's function $G_{\mbox{\tiny E}}(x^{i}, x_0^i)$
        \begin{eqnarray} \label{beq}
        \frac{1}{\sqrt{g^{(3)}}} \frac{\partial}{\partial x^j}
        \left( \sqrt{g^{(3)}} g^{j k} \frac{\partial
        G_{\mbox{\tiny E}}(x^{i},x_0^i)}{\partial x^k}   \right)
        +\frac{g^{j k}}{2 g_{tt}}\frac{\partial g_{tt} }{\partial x^j}
        \frac{\partial G_{\mbox{\tiny E}}
        (x^{i},x_0^i)}{\partial x^k} \nonumber \\
        -\left(m^2+\xi R(x) \frac{}{} \right) G_{\mbox{\tiny E}}(x^{i},x_0^i)
        =-\frac{\delta^{(3)}(x^{i}, x_0^i)}{\sqrt{g^{(3)}}}
        \end{eqnarray}
and introduce the Riemann normal coordinates $y^i$ in 3D space with
origin at the point $x_0^i$ \cite{Pet}. In these coordinates
one has
        \begin{equation}
        g_{i j}(y^i)=\delta_{i j}
        -\frac13 { \widetilde R_{i k j l}}|_{y=0}\, y^k y^l+ O\left(\frac{y^3}{L^3}\right),
        \end{equation}
        \begin{equation}
        g^{i j}(y^i)=\delta^{i j}
        +\frac13 {\widetilde R^{i \ j}_{\ k \ l}}|_{y=0}
        \, y^k y^l+  O\left(\frac{y^3}{L^3}\right),
        \end{equation}
        \begin{equation} \label{exg}
        g^{(3)}(y^i)=1-\frac13 {\widetilde R_{i j}}|_{y=0} y^i y^j
        +  O\left(\frac{y^3}{L^3}\right),
        \end{equation}
where the coefficients here and below are evaluated at $y^i=0$
(i.e. at the point $x_0^i$), and $\delta_{i j}$ denotes the metric
of a flat three-dimensional  Euclidean space-time. $\widetilde
R_{i k j l}$ and $\widetilde R_{i j}$ denote the components of Riemann
and Ricci tensors of the three-dimensional space-time with metric
$g_{i j}$
      \begin{equation} \label{R}
      R_{i j}=\widetilde R_{i j}-\frac{{g_{tt}}_{,i; j}}{2g_{tt}}
      +\frac{{g_{tt}}_{,i} \, {g_{tt}}_{,j}}{4{g_{tt}}^2},
      R=\widetilde R-\frac{{{g_{tt}}_{,i}}^{;i}}{{g_{tt}}}
      +\frac{{g_{tt}}_{,i} \, {g_{tt}}^{,i}}{2{g_{tt}}^2},
      \end{equation}
where ${g_{t t}}_{,{i}}$ denotes the covariant derivative of a
scalar function $g_{t t} (y^j)$ with respect to $y^i$ in 3D space with metric $g_{i j}(y^k)$
(${g_{t t}}_{,{i} ;{j}}$ is the covariant derivative of a vector ${g_{t t}}_{,{i}}$ at point
$y^k=0$ in 3D space, which coincides with the partial derivative as $\Gamma^k_{i j}=0$ at $y^k=0$
in the Riemann normal coordinates).
All indices are raised and lowered with $\delta _{i j}$.
Defining $\overline{G} (y^i)$ by
        \begin{equation} \label{Gw}
        \overline G(y^i)=\sqrt{g^{(3)}} G_{\mbox{\tiny E}}(y^i)
        \end{equation}
and retaining in (\ref{beq}) only the terms with coefficients involving
two derivatives of the metric or fewer one finds that
$\overline G(y^i)$ satisfies the equation
        \begin{eqnarray}
        &&\delta^{i j}\frac{\partial^2 \overline G}{\partial y^i \partial y^j}
        -m^2 \overline G +\delta^{i j} \frac{{g_{tt}}_{,i}}{2 {g_{tt}}}
        \frac{\partial \overline G}{\partial y^j}
        +\delta^{i j}\left(\frac{{g_{tt}}_{,i k}}{2 {g_{tt}}}
        -\frac{{g_{tt}}_{,i} \, {g_{tt}}_{,k}}{2 {g_{tt}}^2}
        \right)y^k \frac{\partial \overline G}{\partial y^j}
        \nonumber \\ &&
         +{{ { {\widetilde R \,} ^i}_{k}}^{j}}_{l} \, \frac{y^k y^l}{3}
        \frac{\partial^2 \overline G}{\partial y^i \partial y^j}
        +\left(\frac{\widetilde R}{3}-\xi R \right)
        \overline G = -\delta^{(3)}(y).
        \end{eqnarray}
Let us present
        \begin{equation} \label{Gy}
        \overline G(y^i)=\overline G_0(y^i) + \overline
        G_1(y^i) + \overline G_2(y^i)+\dots,
        \end{equation}
where $\overline G_a(y^i)$ has a geometrical coefficient
involving $a$ derivatives of the metric at point $y^i=0$. Then
these functions satisfy the equations
        \begin{equation}
        \delta^{i j}\frac{\partial^2 \overline G_0}
        {\partial y^i \partial y^j}-m^2 \overline
        G_0=-\delta^{(3)}(y),
        \end{equation}
        \begin{equation}
        \delta^{i j}\frac{\partial^2 \overline G_1}
        {\partial y^i \partial y^j}-m^2 \overline G_1
        +\delta^{i j}\frac{{g_{tt}}_{,i}}{2{g_{tt}}}\frac{\partial \overline G_0}
        {\partial y^j}=0,
        \end{equation}
        \begin{eqnarray} \label{G2}
        &&\delta^{i j}\frac{\partial^2 \overline G_2}
        {\partial y^i \partial y^j}-m^2 \overline G_2
        +\delta^{i j} \frac{{g_{tt}}_{,i}}{2 {g_{tt}}}
        \frac{\partial \overline G_1}{\partial y^j}
        +\delta^{i j}\left(\frac{{g_{tt}}_{,i k}}{2 {g_{tt}}}
        -\frac{{g_{tt}}_{,i} \, {g_{tt}}_{,k}}{2 {g_{tt}}^2}
        \right)y^k \frac{\partial \overline G_0}{\partial y^j}
        \nonumber \\ &&
         +{{ { {{\widetilde R \,} ^i}}_{k}}^{j}}_{l} \, \frac{y^k y^l}{3}
        \frac{\partial^2\overline  G_0}{\partial y^i \partial y^j}
        +\left(\frac{\widetilde R}{3}-\xi R \right)\overline G_0=0.
        \end{eqnarray}
The function $\overline G_0(y^i)$ satisfies the condition
        \begin{equation}
        {{ { {{\widetilde R \,} ^i}}_{k}}^{j}}_{l} \, y^k y^l  \frac{\partial^2
        \overline G_0}{\partial y^i \partial y^j} -
        {\widetilde R \,} ^i_{j} y^j \frac{\partial \overline G_0}
        {\partial y^i}=0,
        \end{equation}
since $\overline G_0(y^i)$ can be the function only of $\delta_{i j} y^i y^j$.
Therefore Eq. (\ref{G2}) may be rewritten
        \begin{eqnarray} \label{G2n}
        \delta^{i j}\frac{\partial^2 \overline G_2(y^i)}
        {\partial y^i \partial y^j}-m^2 \overline G_2(y^i)
        +\delta^{i j} \frac{{g_{tt}}_{,i}}{2 {g_{tt}}}
        \frac{\partial \overline G_1}{\partial y^j}
        +\left[\frac13 {\widetilde R}^i_{k}+
        \delta^{i j}\left(\frac{{g_{tt}}_{,j k}}{2 {g_{tt}}}
        -\frac{{g_{tt}}_{,j }{g_{tt}}_{,k}}{2 {g_{tt}}^2}
        \right) \right]  \, y^k
        \frac{\partial \overline G_0}{\partial y^i }
        +\left(\frac{\widetilde R}{3} -\xi R\right)\overline G_0=0.
        \end{eqnarray}
Let us introduce the local momentum space associated with the
point $y^i=0$ by making the 3-dimensional Fourier transformation
        \begin{equation} \label{ms}
        \overline G_a(y^i)= \int \!\!\!\!\int \limits_{-\infty}
        ^{+\infty}\!\!\!\!\int \frac{dk_1 dk_2 dk_3}{(2 \pi)^3}
        \exp({i k_i y^i}) \overline G_a(k^i).
        \end{equation}
It is not difficult to see that
        \begin{equation} \label{G0k}
        \overline G_0(k^i)=\frac{1}{k^2+m^2},
        \end{equation}
        \begin{equation} \label{G1k}
        \overline G_1(k^i)=i\frac{\delta^{i j}{g_{tt}}_{,i}k_j}{2{g_{tt}}(k^2+m^2)^2},
        \end{equation}
where $k^2=\delta^{i j}k_i k_j$. In momentum space Eq.(\ref{G2n}) gives
        \begin{equation}
        -(k^2+m^2)\overline G_2(k^i)
        +i\frac{\delta^{i j}{g_{tt}}_{,i}k_j}{2{g_{tt}}}\overline G_1(k^i)
        +\left[ik_i\left(\frac{{\widetilde R}^i_{k}}{3}  +
        \delta^{i j}\frac{{g_{tt}}_{,j k}}{2 {g_{tt}}}
        -\delta^{i j}\frac{{g_{tt}}_{,j} \, {g_{tt}}_{, k}}{2 {g_{tt}}^2}
        \right) \, y^k +\left( \frac{\widetilde R}{3}-\xi R \right) \right]
       \overline G_0(k^i)=0.
        \end{equation}
Hence
        \begin{equation} \label{G2k}
        \overline G_2(k^i)=
        \frac{\displaystyle -\frac{ \delta^{i j}{g_{tt}}_{,i j}}{2 {g_{tt}}}
        +\frac{\delta^{i j}{g_{tt}}_{,i} \, {g_{tt}}_{,j}}{2 {g_{tt}}^2}-\xi R}{(k^2+m^2)^2}
        +\frac{\displaystyle k_i k_j \delta^{i k} \delta^{j l}
        \left( \frac23 {\widetilde R}_{j \, l}
        + \frac{ {g_{tt}}_{,j \, l}}{{g_{tt}}}
        -\frac{5{g_{tt}}_{,j} \, {g_{tt}}_{,l}}{4 {g_{tt}}^2}\right)}{(k^2+m^2)^3}.
        \end{equation}
Substituting (\ref{ms}), (\ref{G0k}), (\ref{G1k}), (\ref{G2k})  in (\ref{Gy})
and integrating leads to
        \begin{eqnarray}
        \overline G_0(y^i) + \overline G_1(y^i) + \overline G_2(y^i)&=&
         \frac{\exp (-m y)}{8 \pi } \left[\frac2y-\frac{{g_{tt}}_{,i}}{2{g_{tt}}} \frac{y^i}{y}
         +\frac{1}{m}\left( -\frac{\delta^{i j}{g_{tt}}_{,i j}}{4 {g_{tt}}}
         +\frac{3\delta^{i j}{g_{tt}}_{,i} \, {g_{tt}}_{,j}}{16 {g_{tt}}^2}-\xi R
         +\frac{\widetilde R}{6}  \right)
         \right. \nonumber \\ && \left.
        +\left( -\frac{{g_{tt}}_{,i j}}{4 {g_{tt}}}+\frac{5{g_{tt}}_{,i} \, {g_{tt}}_{,j}}{16 {g_{tt}}^2}
        -\frac{{\widetilde R}_{i j}}{6} \right) \frac{y^i y^j}{ y}\right],
        \end{eqnarray}
where
        \begin{equation}
        y=\sqrt{\delta_{i j}y^i y^j}.
        \end{equation}
Using the definition of $\overline G(y^i)$ (\ref{Gw}), expansion (\ref{exg}),
and expressions (\ref{R}) one finds
        \begin{eqnarray} \label{Gwy}
        G_{\mbox{\tiny E}}(y^i)&=&\left[1+\frac16 R_{i j}y^i y^j
        +O\left(\frac{y^3}{L^3}\right)\right]\overline G(y^i)
        \nonumber \\ &=&
        \frac{\exp (-m y)}{8 \pi } \left[\frac2y-\frac{{g_{tt}}_{,i}}{2{g_{tt}}} \frac{y^i}{y}
         +\frac{1}{m}\left( -\frac{\delta^{i j}{g_{tt}}_{,ij}}{4 {g_{tt}}}
         +\frac{3\delta^{i j}{g_{tt}}_{,i} \, {g_{tt}}_{,j}}{16 {g_{tt}}^2}-\xi R
         +\frac{{\widetilde R}}{6}  \right)
         \right. \nonumber \\ && \left.
        +\left( -\frac{{g_{tt}}_{,ij}}{4 {g_{tt}}}+\frac{5{g_{tt}}_{,i} \, {g_{tt}}_{,j}}{16 {g_{tt}}^2}
        +\frac{{\widetilde R}_{i j}}{6}  \right) \frac{y^i y^j}{ y}
        +O\left(\frac{1}{m^2 L^3}\right)+O\left(\frac{y}{m L^3}\right)
        +O\left(\frac{y^2}{L^3}\right)\right]
         \nonumber \\ &=&
         \frac{1}{8 \pi } \left\{\frac2y-\frac{{g_{tt}}_{,i}}{2{g_{tt}}}\frac{y^i}{y}-2m
         +\frac{1}{m}\left[ -\frac{\delta^{i j}{g_{tt}}_{,i j}}{12 {g_{tt}}}
         +\frac{5\delta^{i j}{g_{tt}}_{,i} \, {g_{tt}}_{,j}}{48 {g_{tt}}^2}-\left(\xi -\frac16 \right) R \right]
         \right. \nonumber \\ && \left.
         +m\frac{{g_{tt}}_{,i} \, y^i}{2{g_{tt}} }+m^2 y
         +\left[ \frac{\delta^{i j}{g_{tt}}_{,i j}}{12 {g_{tt}}}
         -\frac{5\delta^{i j}{g_{tt}}_{,i} \, {g_{tt}}_{,j}}{48 {g_{tt}}^2}+\left(\xi -\frac16 \right) R \right]y
         \right. \nonumber \\ && \left.
        +\left( -\frac{{g_{tt}}_{,i j}}{6 {g_{tt}}}+\frac{13{g_{tt}}_{,i} \, {g_{tt}}_{,j}}{48 {g_{tt}}^2}
        +\frac{R_{i j}}{6}  \right) \frac{y^i y^j}{ y}
        \right. \nonumber \\ && \left.
        +O\left(\frac{1}{m^2 L^3}\right)+O\left(\frac{y}{m L^3}\right)
        +O\left(\frac{y^2}{L^3}\right) +O\left(\frac{m y^3}{L^3}\right)
        +O\left(m^3 y^2\right)\right\}.
        \end{eqnarray}
In the arbitrary coordinates of 3D space
    \begin{equation}
    y^i \rightarrow u^{i}(x_0) \Delta s \equiv -\sigma^{i},
    \end{equation}
where $u^{i}(x_0)$ is the unit tangent vector to the shortest geodesic connecting
points $x_0$ and $x$ which is calculated at points $x_0$ and directed from $x_0$
to $x$, $\Delta s$ is the distance between these points along the considered
geodesic. Therefore
        \begin{eqnarray} \label{Gwsigma}
        G_{\mbox{\tiny E}}(x^i; x_0^i)
        &=&
        \frac{1}{8 \pi } \left\{\frac{2}{\sqrt{2 \sigma}}
        +\frac{{{g}_{t t}}_{,{i}}\sigma^{{i}}}
        {2{{g}_{t t}} \sqrt{2 \sigma}}-2m
         +\frac{1}{m}\left[ -\frac{{{{g}_{t t}}_{,{i}}}^{;{i}}}
         {12 {{g}_{t t}}}
         +\frac{5{{g}_{t t}}_{,{i}}{{g}_{t t}}^{,{i}}}
         {48 {{g}_{t t}}^2}
         -\left(\xi -\frac16 \right) R(x_0) \right]
         \right. \nonumber \\ && \left.
         -m \frac{{{g}_{t t}}_{,{i}}\sigma^{{i}}}{2{{g}_{t t}} }
         +m^2 \sqrt{2 \sigma}
         +\left[ \frac{{{{g}_{t t}}_{,{i}}}^{;{i}}}
         {12 {{g}_{t t}}}
         -\frac{5{{g}_{t t}}_{,{i}} \, {{g}_{t t}}^{,{i}}}
         {48 {{g}_{t t}}^2}
         +\left(\xi -\frac16 \right) R(x_0) \right]\sqrt{2 \sigma}
         \right. \nonumber \\ && \left.
        +\left( -\frac{{{g}_{t t}}_{,{i} ;{j}}}{6 {{g}_{t t}}}
        +\frac{13{{g}_{t t}}_{,{i}} \,
        {{g}_{t t}}_{,{j}}}{48 {{g}_{t t}}^2}
        +\frac{{R}_{{i} {j}}(x_0)}{6}  \right)
        \frac{\sigma^{{i}} \sigma^{{j}}}{ \sqrt{2 \sigma}}
        \right. \nonumber \\ && \left.
        +O\left(\frac{1}{m^2 L^3}\right)+O\left(\frac{\sqrt{\sigma}}{m L^3}\right)
        +O\left(\frac{\sigma}{L^3}\right) +O\left(\frac{m \sigma^{3/2}}{L^3}\right)
        +O\left(m^2 \sigma \right) \right\},
        \end{eqnarray}
where ${g_{t t}}_{,{i}}$ denotes the covariant derivative of a
scalar function $g_{t t} (x_0)$ with respect to $x_0^i$ in 3D space with metric $g_{i j}(x_0)$
(${g_{t t}}_{,{i} ;{j}}$ is the covariant derivative of a vector ${g_{t t}}_{,{i}}$ at point
$x_0$ in 3D space),
\begin{equation}
\sigma= \frac{g_{i j }(x_0)}{2}  {\sigma^i} {\sigma^j}
\end{equation}
is one-half the square of the distance
between the points $x_0^i$ and $x^i$ along the shortest geodesic connecting them, and
(see, e.g., \cite{Synge:1960,Popov:2007})
        \begin{eqnarray}
        {\sigma^i}&=&-\left(x^i-x_0^i\right)
        -\frac12 \Gamma^{i}_{{j}{k}}\left(x^j-{x_0^j}\right)\left(x^k-{x_0^k}\right)
        \nonumber \\ &&
        -\frac16 \left( \Gamma^{i}_{{j}{m}} \Gamma^{m}_{{k}{l}}
        +\frac{\partial \Gamma^{i}_{{j}{k}}}{\partial {x_0^l}}\right)
        \left(x^j-{x_0^j}\right)\left(x^k-{x_0^k}\right)\left(x^l-{x_0^l}\right)
        +O\left(\left(x-{x_0}\right)^4\right),
        \end{eqnarray}
where the Christoffel symbols $\Gamma^{i}_{{j}{k}}$ are calculated at the point $x_0$.

Now we can use the expansion of
\begin{equation}
\phi_m(x^i; x_0^i)=4 \pi q G_{\mbox{\tiny E}}(x^i; x_0^i)
\end{equation}
in the regularization procedure (\ref{Rosenthal}). But if we take the limits before
the partial differentiation in (\ref{Rosenthal}), then the last two terms do not appear
in the expression for $f_\mu^{self}(x_0)$. And in the considered case of a charge at rest
in a static space-time we can renormalize the self-potential as
\begin{equation} \label{ren}
\phi_{ren}(x)=\lim_{x_0 \rightarrow x}\left( \phi(x; x_0)
-\phi_{\mbox{\tiny DS}}(x; x_0) \right),
\end{equation}
where
        \begin{equation} \label{phiDS}
        \phi_{\mbox{\tiny DS}}(x^i; x_0^i)=
        q \left(\frac{1}{\sqrt{2 \sigma}}
        +\frac{\partial g_{t t}(x_0)}{ \partial x^i_0}
        \frac{\sigma^{{i}}}{4g_{t t}(x_0)\sqrt{2 \sigma}}-m
        \right),
        \end{equation}
and $\phi(x; x_0)$ is the solution of (\ref{meq}) in the case of arbitrary
mass $m$ (even $m=0$).
Finally the self-force acting on a static scalar charge is
\begin{equation} \label{sf}
f_\mu^{self}(x)=-\frac{q}{2}\frac{\partial \phi_{ren}(x)}{\partial x^\mu}.
\end{equation}

\section{The Schwarzschild space-time}\label{Sec:BlackHole}

Let us verify the above scheme for the well-known case of a black hole space-time
\cite{Lin77, Wise:2000}
\begin{equation}
ds^2 = - f(r) dt^2 + \frac{dr^2}{f(r)} + r^2 (d\theta^2 + \sin^2\theta d\varphi^2),
\quad f(r)=\left(1 - \frac{2M}r\right).
\end{equation}
In the case $x^i-x_0^i=\delta^i_r (r-r_0)$
\begin{eqnarray}
\sigma^{r}&=&-(r-r_0)+\frac{1}{4 f(r_0) }\frac{d f(r_0)}{dr_0} (r-r_0)^2
\nonumber \\ &&
-\frac16 \left( \frac{3}{4 f(r_0)^2} \left( \frac{d f(r_0)}{dr_0} \right)^2
-\frac{\qquad d^2 f(r_0)}{2 f(r_0) dr_0^2}\right)(r-r_0)^3
+O\left( \frac{(r-r_0)^4}{L^3} \right),
\end{eqnarray}
\begin{eqnarray}
\sigma&=&\frac{(r-r_0)^2}{2 f(r_0)}\left[1-\frac{\qquad d f(r_0)}{2 f(r_0) dr_0} (r-r_0)
\right. \nonumber \\ && \left.
+\left( \frac{5}{16 f(r_0)^2} \left( \frac{d f(r_0)}{dr_0} \right)^2
-\frac{\qquad d^2 f(r_0)}{6 f(r_0) dr_0^2} \right)(r-r_0)^2\right]
+O\left( \frac{(r-r_0)^5}{L^3} \right),
\end{eqnarray}
and the expansion of $\phi(r; r_0)$ for the massive field ($mL \gg 1$) is
\begin{eqnarray} \label{phiS}
\phi_m(r; r_0)&=&4\pi q G_{\mbox{\tiny E}}(r; r_0)
=\frac{\displaystyle q \sqrt{1- \frac{\displaystyle 2M}{\displaystyle r_0}} }{|r-r_0|} -q m
+\frac{q}{2} \left( -\frac{M}{\displaystyle 3 {r_0}^3 \sqrt{1-\frac{2M}{r_0}}}
\right. \nonumber \\ && \left.
+\frac{m M}{\displaystyle {r_0}^2 \left(1-\frac{2M}{r_0}\right)}
+\frac{m^2}{\displaystyle \sqrt{1-\frac{2M}{r_0}}}
\right)(r-r_0)+O\left( \frac{q}{mL^2} \right)+O\left( \frac{q(r-r_0)}{mL^3} \right)
\nonumber \\ &&
+O\left(\frac{q(r-r_0)^2}{L^3} \right)+O\left(q m^2(r-r_0)^2 \right)
+O\left(\frac{q m(r-r_0)^3}{L^3} \right).
\end{eqnarray}
Therefore the renormalized counterterm for the massless field is
\begin{equation} \label{phish}
\phi_{\mbox{\tiny DS}}(r; r_0) = \frac{q \sqrt{f(r_0)}}{|r-r_0|}.
\end{equation}
This expression coincides with  the unrenormalized potential $\phi(x; x')$ of
a scalar point charge at rest in Schwarzschild space-time
in the case $\xi=0$, $m=0$ and $t=t', \theta=\theta', \varphi=\varphi'$ \cite{Lin77}.
Consequently the renormalized expression for the self-potential is
\begin{equation}
\phi_{ren}(r)=0, \quad \mbox{and} \quad f_i^{self}(r)=0.
\end{equation}
Note the derivative of (\ref{phiS}) ${\partial\phi(r; r_0) }/{\partial r}$ differs from
the corresponding expression in \cite{2Roth:2004} by the term
$-qM/(6 {r_0}^3\sqrt{1-2M/r_0})$.

\section{Conclusion} \label{Sec:Conclusion}

The considered approach gives the possibility to renormalize (\ref{ren})
the self-potential of scalar point charge $q$ at rest in static space-time (\ref{metric})
and to calculate the self-force (\ref{sf}) acting on this charge.
Note that in the case in which the Compton wavelength $1/m$ of the massive scalar field is much smaller
than the characteristic scale $L$ of curvature of the background gravitational field
at the considered point $x$ we can obtain the approximated expression
for the renormalized self-potential
\begin{eqnarray} \label{fin}
         \phi_{ren}(x)&=&\lim_{x_0 \rightarrow x}\left( \phi_m(x; x_0)
         -\phi_{\mbox{\tiny DS}}(x; x_0) \right)
         \nonumber \\ &&
         =\frac{q}{2 m}\left[ -\frac{{{g_{t t}}_{,{i}}}^{;{i}}}{12 {g_{t t}}}
         +\frac{5{g_{t t}}_{,{i}}{g_{t t }}^{,{i}}}{48 {g_{t t}}^2}
         -\left(\xi -\frac16 \right) R \right]
         +O\left( \frac{q}{m^2 L^3} \right).
\end{eqnarray}
Of course the order of this expression in $1/(mL)$ is less than the correspondent
order of $\phi_{ren}$ for the massless field (or field with mass
$m \mathop{\lefteqn{\raise.9pt\hbox{$<$}}\raise-3.7pt\hbox{$\sim$}} 1/L$).
However the expression (\ref{fin}) can be used for the verification of asymptotic behavior
of $\phi_{ren}$ in the limit $m \rightarrow \infty$.

\section*{Acknowledgments}

This work was supported in part by Grants No. 11-02-01162 and
No. 11-02-90477 from the Russian Foundation for Basic Research.


\end{document}